\documentclass[a4paper, amsfonts, amssymb, amsmath, reprint, showkeys, nofootinbib]{revtex4-1}
\usepackage[english]{babel}
\usepackage[utf8]{inputenc}
\usepackage[colorinlistoftodos, color=green!40, prependcaption]{todonotes}
\usepackage{amsthm}
\usepackage{mathtools}
\usepackage{physics}
\usepackage{xcolor}
\usepackage{graphicx}
\usepackage[left=23mm,right=13mm,top=35mm,columnsep=15pt]{geometry} 
\usepackage{adjustbox}
\usepackage{placeins}
\usepackage[T1]{fontenc}
\usepackage{lipsum}
\usepackage{csquotes}

\usepackage{multirow}
\usepackage[pdftex, pdftitle={Article}, pdfauthor={Author}]{hyperref} 
\hypersetup{
  colorlinks     = true, 
  linkbordercolor= blue,
  urlcolor       = blue, 
  linkcolor      = blue, 
  citecolor      = blue 
}
\definecolor{lime}{HTML}{A6CE39}
\definecolor{dred}{rgb}{0.8,0,0.1}
\definecolor{orange}{rgb}{1,0.5,0}

\DeclareRobustCommand{\orcidicon}{\hspace{-4pt}
\begin{tikzpicture}
\draw[lime, fill=lime] (0,0) 
circle [radius=0.16] 
node[white] {\hspace{0.1mm}{\fontfamily{qag}\selectfont \tiny ID}};
\draw[white, fill=white] (-0.07,0.1) 
circle [radius=0.01];
\end{tikzpicture}
\hspace{-3.2mm}
}

\foreach \x in {A, ..., Z}{\expandafter\xdef\csname 
orcid\x\endcsname{\noexpand\href{https://orcid.org/\csname orcidauthor\x\endcsname}
{\noexpand\orcidicon}}
}

\renewcommand{\arraystretch}{1.2}
\begin{document}

\title{Exploring cosmological imprints of phantom crossing with dynamical dark
energy in Horndeski gravity}

\author{$^1$Yashi Tiwari\orcidA{}}
\email{yashitiwari@iisc.ac.in}

\author{$^{1,2}$Ujjwal Upadhyay\orcidB{}}
    \email{ujjwalu@iisc.ac.in}

\author{$^3$Rajeev Kumar Jain\orcidC{}}
    \email{rkjain@iisc.ac.in}
    
    \affiliation{$^1$Joint Astronomy Programme, Department of Physics, Indian Institute of Science, C. V. Raman Road, Bangalore 560012, India}

    \affiliation{$^2$Astronomy \& Astrophysics Group, Raman Research Institute, C. V. Raman Avenue, Bangalore 560080, India}
    
    \affiliation{$^3$Department of Physics, Indian Institute of Science, C. V. Raman Road, Bangalore 560012, India\\}

\begin{abstract}
In the current era of precision cosmology, the persistence of cosmological tensions—most notably the Hubble tension and the $S_8$ tension—challenges the standard $\Lambda$CDM model. To reconcile these tensions via late-time modifications to expansion history, various features such as phantom crossing in the dark energy equation of state, a negative energy density at high redshifts, etc., are favoured. However, these scenarios cannot be realized within the framework of GR without introducing ghost or gradient instabilities. In this work, we investigate a dynamical dark energy scenario within the framework of Horndeski gravity, incorporating nonminimal coupling to gravity and self-interactions. We highlight that the model can exhibit novel features like phantom crossing and negative dark energy densities at high redshifts without introducing any instabilities. For this specific Horndeski model, we perform a comprehensive analysis of the background evolution along with the effects on perturbations, examining observables like growth rate, matter and CMB power spectrum. To check the consistency of the model with the observational data, we employ MCMC analysis using BAO/$f\sigma_8$, Supernovae, and CMB data. While the model does not outperform the standard $\Lambda$CDM framework in a combined likelihood analysis, there remains a preference for non-zero values of the model parameters within the data. This suggests that dynamical dark energy scenarios, particularly those with non-minimal couplings, merit further exploration as promising alternatives to GR, offering rich phenomenology that can be tested against a broader range of current and upcoming observational datasets.

\end{abstract}


\maketitle

\section{Introduction} \label{sec:first}
Advances in observational techniques and statistical inference have ushered cosmology into a precision era, enabling constraints on cosmological parameters at the sub-percent level. While this progress has strengthened confidence in the standard Lambda-Cold Dark Matter ($\Lambda$CDM) cosmology, the past decade has also revealed inconsistencies among certain cosmological parameters inferred from different datasets. The most prominent of these are the Hubble tension and the growth tension.
Hubble tension refers to the significant discrepancy between the Hubble constant, $H_0$, 
measured from local universe observations, such as distance ladder calibrations, and the value inferred from Cosmic Microwave Background (CMB) observations under the assumption of the $\Lambda$CDM model.  For instance, the SH0ES collaboration \cite{Riess:2021jrx} reports a value of $H_0= 73.04 \pm 1.04$ km s$^{-1}$ Mpc$^{-1}$, while the Planck best-fit $\Lambda$CDM value  is $H_0= 67.4 \pm 0.5$ km s$^{-1}$ Mpc$^{-1}$, resulting in a tension of $5\sigma$ significance \cite{Planck:2018vyg}. The growth tension pertains to discrepancies in the late-time cosmic structure formation parameters $\sigma_8$ (the amplitude of matter density fluctuations averaged over a scale of $8h^{-1}$ Mpc)  and $S_8 \equiv \sigma_8\sqrt{\Omega_m/0.3}$, where $\Omega_m$ is the present value of the matter density parameter. Weak lensing surveys like the Kilo-Degree Survey (KiDS) \cite{KiDS:2020suj} report $S_8=0.759 \pm 0.021$, which is lower than the CMB-inferred value of $S_8=0.834\pm 0.016$, leading to a $2-3\sigma$ tension. While this level of tension might seem modest, it can introduce significant uncertainty into cosmological simulations of large-scale structures. Moreover, the growth tension can also be linked to deviations from General Relativity (GR) when the growth rate of matter perturbations is parameterized using the 'growth index' $\gamma$. The growth rate is approximated as $f(a)=\Omega_m(a)^\gamma$, where $\gamma = 0.55$ is the GR prediction for a flat $\Lambda$CDM model. However, as highlighted in \cite{Nguyen:2023fip}, a model-independent analysis incorporating $f\sigma_8$ measurements from peculiar velocities and Redshift Space Distortions (RSD), combined with Planck 2018 data, finds $\gamma = 0.639^{+0.025}_{-0.024}$, a $3.7\sigma$ deviation from the GR expectation. This elevated $\gamma$ suggests a suppressed growth of structures during the dark energy-dominated era, aligning with the lower $S_8$ value inferred from weak lensing measurement and hinting at a possible deviation from GR. 
Assuming that these tensions are not due to some unaccounted observational systematics in the analysis, several approaches have been proposed to resolve them by modifying the standard $\Lambda$CDM model. In the case of the Hubble tension, solutions fall into two broad categories: modifying the physics before recombination to reduce the sound horizon or introducing late-time modifications to the expansion history. Both approaches aim to preserve the angular size of the sound horizon at the time of recombination, $\theta_*$, a precisely measured cosmological parameter from CMB observations (See, for instance, Refs. \cite{Knox:2019rjx, Schoneberg:2021qvd, DiValentino:2021izs, Abdalla:2022yfr, Hu:2023jqc} for a comprehensive review). However, most of the solutions that attempt to resolve the Hubble tension exacerbate the growth tension. In Refs. \cite{Jedamzik:2020zmd, Vagnozzi:2023nrq}, it has been demonstrated that early-universe solutions alone are insufficient to alleviate both tensions, often requiring a combination of early- and late-time modifications (see \cite{Simon:2024jmu}, for example). Meanwhile, simple late-time modifications, such as a phantom dark energy equation of state, struggle to reconcile with Baryon Acoustic Oscillation (BAO) measurements and Supernova data,  which tightly constrain the late-time expansion history \cite{Camarena:2021jlr, Yang:2021hxg}. It has also been pointed out in \cite{Heisenberg:2022lob, Tutusaus:2023cms, Lee:2022cyh}  that for late-time modifications to resolve cosmological tensions without disturbing the CMB measurements, it requires a phantom crossing, where the dark energy equation must transit between phantom and non-phantom regimes at some redshift.  As a result, the simultaneous resolution of both the $H_0$ and $S_8$ tensions remains a complex and challenging problem. 
\par In recent years, with the influx of low-redshift data, parametric and non-parametric reconstructions of the expansion history and dark energy properties, such as energy density and the equation of state, have been pursued to better understand cosmological tensions and the nature of dark energy. Many of these studies have pointed out that the late time expansion history may not be as trivial as explained by the cosmological constant but rather could be more dynamical in nature with interesting features like phantom crossing or the negative dark energy density at high redshifts \cite{Wang:2018fng, Dutta:2018vmq, Escamilla:2021uoj, Akarsu:2022lhx, Malekjani:2023ple, Gomez-Valent:2023uof, Sabogal:2024qxs}. The concept of negative dark energy at high redshifts was initially explored as a potential explanation for the anomalous Ly-$\alpha$ measurements of the expansion history by eBOSS around $z \sim 2.33$, which could not be explained by the $\Lambda$CDM model \cite{eBOSS:2020tmo, BOSS:2014hhw, Sahni:2014ooa, Manoharan:2024thb}.  Recent studies have examined scenarios wherein dark energy density or equation of state is parametrized to accommodate these features, and they are found to give a good fit to the observational data, reconciling cosmological tensions \cite{Akarsu:2019hmw, Teng:2021cvy, Sen:2021wld, Akarsu:2022typ, Adil:2023exv, Dwivedi:2024okk}. Interestingly, the first data release of BAO measurements by Dark Energy Spectroscopic Instrument (DESI) has also hinted towards a dynamical nature of dark energy when combined with CMB and Supernovae data \cite{DESI:2024mwx, DESI:2024uvr, DESI:2024aqx, DESI:2024kob}.
This analysis is conducted for a CPL parameterization of the dark energy equation of state \cite{Chevallier:2000qy}, i.e., specifically within the $w_0w_a$CDM scenario. Subsequent reconstructions of the equation of state, using DESI, CMB, and SNIa data, have confirmed that this dynamical behaviour is not merely a consequence of the chosen CPL parameterization but is instead a feature supported by the data itself \cite{Ye:2024ywg, Giare:2024gpk, Mukherjee:2024ryz, Chudaykin:2024gol}. Notably, these studies have also uncovered hints of a phantom crossing at low redshifts, where the equation of state transitions to $w<-1$ around $z < 1$. This makes it an intriguing direction to explore and revive our understanding of dark energy.  If future observations continue to support evidence of its dynamical nature, along with the possibility of phantom behavior at certain redshifts, such phenomena would not be explained within the GR framework without invoking instabilities \cite{Vikman:2004dc, Deffayet:2010qz, Wolf:2024stt}. Henceforth, it calls for the need to go beyond GR. Horndeski theory, a generalized scalar-tensor theory of gravity, is an excellent avenue to explore the simplest extension of GR, offering a rich phenomenology \cite{Horndeski:1974wa, Kobayashi:2019hrl, Bellini:2014fua}. With a Lagrangian constructed from a scalar field and metric tensor, the theory gives second-order equations of motion. In fact, Horndeski theory can accommodate various subclasses of Modified Gravity (MG) models, including Brans-Dicke, Galileon, Quintessence, K-essence, and even interacting dark sectors \cite{Bansal:2024bbb} (such as chameleons), etc., along with GR as well. What makes Horndeski theory interesting is its phenomenological framework for exploring dynamical dark energy scenarios, particularly in light of emerging cosmological tensions. This is because it can accommodate novel features like phantom divide without ghost or gradient stabilities, as shown in various studies, which have become even more intriguing after the DESI data release \cite{Matsumoto:2017qil, Ye:2024ywg, Chudaykin:2024gol}.
\par In this work, we investigate a specific class of Horndeski theory featuring a dynamical dark energy scalar field with nonminimal coupling to gravity and self-interactions. We demonstrate that this model, free from various instabilities, exhibits intriguing dynamics, such as crossing the phantom divide and negative dark energy density at high redshifts. This study builds on our previous work \cite{Tiwari:2023jle}, where we established that this class of models possesses the necessary properties to address cosmological tensions, particularly the $H_0$ tension. We showed that the model alleviates the $H_0$ tension to $2.5\sigma$ when confronted with Supernovae, BAO, and Cosmic Chronometers (CC) data. In this paper, we extend the analysis to explore the effects on the growth of perturbations, including the CMB and matter power spectra, driven by the novel features of this dynamical dark energy scenario. To provide a comprehensive assessment, we explore the parameter space of the model using the Markov Chain Monte Carlo (MCMC) sampling technique incorporating data from Supernovae, BAO/$f\sigma_8$, and CMB measurements. 
\par This paper is organized as follows. In Sec. \ref{sec:second}, we describe the framework of Horndeski gravity and the motivation for its use in the present work. In Sec. \ref{sec:third}, we present our model and discuss the mathematical formulation of background and perturbation dynamics in terms of the $G_i$ functions of Horndeski theory. In Sec. \ref{sec:fourth}, we discuss the non-trivial features of the model and how it depends on the strength of different coupling parameters. In Sec. \ref{sec:fifth}, we describe the data used in the MCMC analysis of the model and present the results of the analysis. We conclude in Sec. \ref{sec:sixth} with a discussion of the results. In Appendix \ref{sec:appendix}, we present the general expressions for the background evolution of a dark energy model within the Horndeski theory. We also discuss the necessary stability conditions in the evolution of perturbations for a consistent theory.  
\section{Horndeski Gravity} \label{sec:second}
Horndeski gravity, a generalized scalar-tensor theory, is an extension of GR by introducing an additional degree of freedom in the form of a scalar field. The Lagrangian is constructed using the metric tensor and a scalar field in four dimensions while ensuring that the equations of motion remain second-order, thereby avoiding Ostrogradsky instabilities \cite{Motohashi:2014opa}. The Lagrangian for Horndeski gravity is given as:
\begin{equation}
\mathcal{L}= \sum_{i=2}^5 \mathcal{L}_{i}\,,
\label{e: horndeski_lag}
\end{equation}

where
\begin{eqnarray}
     && \mathcal{L}_2\ =\  G_2(\phi, X),\nonumber\\
     && \mathcal{L}_3\ =\  -G_3(\phi, X)\Box{\phi},\nonumber\\
     && \mathcal{L}_4\ =\  G_4(\phi, X)R + G_{4,X}(\phi, X)\Big[(\Box{\phi})^2 - (\nabla_{\mu}\nabla_{\nu}\phi)^2\Big],\nonumber\\
      &&\mathcal{L}_5\ =\  G_5(\phi,X)G_{\mu\nu}\nabla^{\mu}\nabla^{\nu}\phi - \frac{1}{6}  G_{5,X}(\phi, X)\nonumber \\ &&\;\;\;\;\;\;\;\;\;\;\;\times\Big[(\Box{\phi})^3-  
      3\Box{\phi}(\nabla_{\mu}\nabla_{\nu}\phi)^2+2(\nabla_{\mu}\nabla_{\nu}\phi)^3\Big], \nonumber
\end{eqnarray}
where $R$ is the Ricci scalar, $G_i$ are four independent arbitrary functions of $\phi$ and $X$, and $X=\partial_{\mu}\phi\partial^{\mu}\phi$/2, $G_{i,Y} =\partial{G}_i/\partial{Y}$ with $Y=\{\phi,X\}$. 
%
Thus, in the Horndeski gravity, the complete action can be given as,
\begin{equation}
\mathcal{S} = \int \mathrm{d}^4x \sqrt{-g}\, \left(\sum_{i=2}^5 \mathcal{L}_{i}+\mathcal{L}_M\right)\,,
\label{e:horndeski_action}
\end{equation}
where,
$\mathcal{L}_M$ accounts for the matter and the radiation components. In a flat FLRW background, the Friedmann equations and the evolution of the scalar field are derived by varying the Horndeski action with respect to the metric tensor and the scalar field \cite{Kobayashi:2011nu, Kobayashi:2019hrl}, the expressions of which are provided in Appendix \ref{sec:appendix}. The flexibility in choosing the functional form of the coupling functions ($G_i$) in the Horndeski Lagrangian makes model building within this framework particularly versatile, allowing for a wide range of scalar-tensor models. For scenarios where the scalar field behaves as dark energy and drives the expansion of the universe, the evolution of cosmological perturbations has been thoroughly studied in \cite{Bellini:2014fua}. Following this, the dynamics of Horndeski gravity has been implemented in the {\tt hi\_class} code \cite{Zumalacarregui:2016pph, Bellini:2019syt}, an extension of the well-known Boltzmann code {\tt class} \cite{Blas:2011rf}. The {\tt hi\_class} code has been developed to handle Horndeski models in a highly generic and flexible manner, making it a powerful tool for exploring a wide array of dark energy and modified gravity scenarios, particularly when investigating their impact on the CMB and large-scale structure.
\par When exploring dark energy scenarios, a critical concern is the potential for instabilities, particularly ghost and gradient instabilities, which can emerge in complex non-canonical theories like Horndeski gravity. Ghost instabilities occur when the kinetic term of perturbations has the wrong sign, leading to unphysical runaway energy growth \cite{Clifton:2011jh}. Gradient instabilities, on the other hand, arise when the squared sound speed of scalar perturbations becomes negative, causing rapid, uncontrollable growth of perturbations on small scales \cite{Hsu:2004vr, Quiros:2017gsu}. Both types of instability significantly constrain the physical viability of scalar-tensor models within the Horndeski framework. The specific conditions required to avoid ghost and gradient instabilities for scalar perturbations are outlined in Appendix \ref{sec:appendix}. Ensuring that a dark energy scenario remains stable and free of such pathologies is essential to its physical meaningfulness, allowing for robust exploration of rich dynamics without encountering instabilities. Notably, the {\tt hi\_class} code integrates the dynamics of Horndeski gravity in a way that automatically checks for and eliminates these pathologies by enforcing appropriate stability conditions, ensuring that the model remains both stable and viable.

\section{The model} \label{sec:third}
In this section, we introduce our dark energy scenario within the framework of Horndeski gravity, where the cosmological constant is replaced by a dynamical scalar field. This construction allows for a richer phenomenology, as will be discussed in the later sections. The complete Lagrangian for our model is given by

\begin{eqnarray}
    \mathcal{L}_\phi &=& \frac{1}{2} \partial_{\mu}\phi \partial^{\mu}\phi - V(\phi)
    - \left[ c_1 \phi + \frac{1}{2} c_2 \partial_{\mu}\phi \partial^{\mu}\phi \right] 
    \partial_{\mu} \partial^{\mu} \phi \nonumber\\
    && + R \left[ \frac{1}2 + c_3 \phi \right],
\label{e:lagrangian}
\end{eqnarray}
where $\phi$ is the scalar field representing dark energy, $V(\phi)$ is the potential, and $c_1$, $c_2$, and $c_3$ are coupling constants that determine the strength of various interactions of the scalar field. This model can be understood as a subclass of Horndeski gravity by identifying the functions $G_i$ as:
\begin{eqnarray}
    G_2=X-V(\phi),\quad G_3=c_1\phi+c_2 X, \nonumber \\
    G_4=\frac{1}{2}+c_3\phi,\quad \quad \quad \quad G_5=0\quad. 
\label{e:setup}
\end{eqnarray}
where $X = \frac{1}{2} \partial_{\mu} \phi \partial^{\mu} \phi$ is the kinetic term of the scalar field.
The function $G_2$ represents the standard kinetic and potential terms for the scalar field, similar to that in a canonical scalar field theory. When $G_3 = 0$ and $G_4 = \frac{1}{2}$, i.e. $c_1=c_2=c_3=0$, this model reduces to the canonical Quintessence scenario, which mimics the dynamics of the $\Lambda$CDM model but with a time-varying dark energy component instead of a cosmological constant. The $G_3$ term introduces non-trivial self-interactions of the scalar field, which are crucial for modifying the cosmological evolution, particularly at late times. The $G_4$ term, on the other hand, represents a non-minimal coupling between the scalar field and gravity, which can affect the strength of gravity, influencing both the background expansion and the growth of cosmic structures.
Finally, we set $G_5 = 0$ to avoid superluminal propagation of tensor modes, i.e., gravitational waves (GWs). This is motivated by stringent constraints from the observation of GWs and their electromagnetic counterparts by the LIGO-VIRGO collaboration \cite{LIGOScientific:2017zic}. These observations imply that the speed of GWs must be extremely close to the speed of light, severely limiting the functional form of $G_5$ in Horndeski theories, as discussed in \cite{Creminelli:2017sry, Gong:2017kim, Kase:2018aps}. In particular, it is the inclusion of scalar field interactions in the form of $G_3$ and the non-minimal coupling in $G_4$ that introduces significant deviations from the standard cosmological evolution, especially at late times. In addition, these terms also modify the growth of structures, as will be discussed in later sections.
\par To investigate the evolution of both the background and perturbations, we implement our model in the {\tt hi\_class} code. The governing equations for the background dynamics—including the Friedmann equation, the scalar field evolution, and its corresponding energy density and pressure—are derived from the general expressions provided in Appendix \ref{sec:appendix}, where we substitute Eq. (4) to specialize them for our scenario. In particular, we adopt a linear potential for the scalar field, $V(\phi)=V_0 \phi$.
The {\tt hi\_class} code evolves the background equations by using initial conditions on Hubble parameter, scalar field and its derivative while $V_0$ is used as a tuning parameter. This parameter is calibrated via the shooting method to ensure the correct present-day value of the dark energy density. The initial conditions for the scalar field and its derivative are set shortly after recombination at $z\sim1000$ with values $\phi_i=10.0$ and $\phi^{'}_i=10^{-10}$ in Planck units where $\phi^{'}=d\phi/d\tau$, where $\tau$ is the conformal time. Consequently, the model contains three free parameters— $c_1$,$c_2$ and $c_3$. The choice of initial conditions on $\phi$ and $\phi^{'}$ is motivated such that when the model parameters are set to zero ($c_1=c_2=c_3=0$), the system reverts to canonical Quintessence, which is similar to $\Lambda$CDM universe, ensuring consistency with the standard cosmological model as a baseline.  Varying the free parameters reveals rich and interesting features in the dynamics of the scalar field and the expansion history of the universe.
In addition to the background evolution, {\tt hi\_class} computes the evolution of key cosmological observables, such as the growth rate, the matter power spectrum, and the CMB power spectrum. This allows for a comprehensive exploration of how different parameter choices impact both large-scale structure formation and the CMB anisotropies.
Finally, to ensure a consistent and physically viable model, the evolution is kept free from ghost and gradient instabilities by using appropriate configurations within the {\tt hi\_class} code.
In \texttt{hi\_class}, a mix of natural units (\(\hbar = c = 1\)) and cosmological conventions (expressing all quantities in terms of \(\mathrm{Mpc}\)) is used, with the scalar field normalized to be dimensionless. In this framework, \(c_1\) and \(c_3\) are dimensionless parameters, while \(c_2\) is expressed in units of \(\mathrm{Mpc}^2\).

\section{Non-trivial Dynamics}\label{sec:fourth}
The presence of non-minimal coupling and self-interactions has invoked certain non-trivial dynamics in the model, making it interesting to explore as a candidate for dark energy. Unlike the cosmological constant in the standard $\Lambda$CDM model, which remains static over cosmic time, dark energy in this framework is dynamical, characterized by an evolving equation of state parameter, $w_\phi$. One of the distinguishing features of this model is its tendency to enter the so-called phantom regime, where $w_\phi < -1$, during certain epochs of cosmic evolution. This involves the notable phenomenon of phantom crossing, where the equation of state makes a transition across the phantom divide line, $w = -1$. This is an important aspect of the model because phantom behaviour is often linked to ghost instabilities in conventional GR frameworks. However, in this case, it occurs without such instabilities. 
Furthermore, the model exhibits another intriguing feature: a negative energy density of the scalar field at high redshifts. As discussed in Sec. \ref{sec:first}, the negative energy density at early times aligns with various cosmological observations, offering a potentially better fit than models with a constant or positive dark energy density. The rich and complex dynamics of the model, achieving phantom crossing and negative energy densities without invoking instabilities, make it a compelling alternative to $\Lambda$CDM and conventional scalar field models.
\subsection{Background}
\begin{figure}
    \centering
    \includegraphics[width=0.52\textwidth]{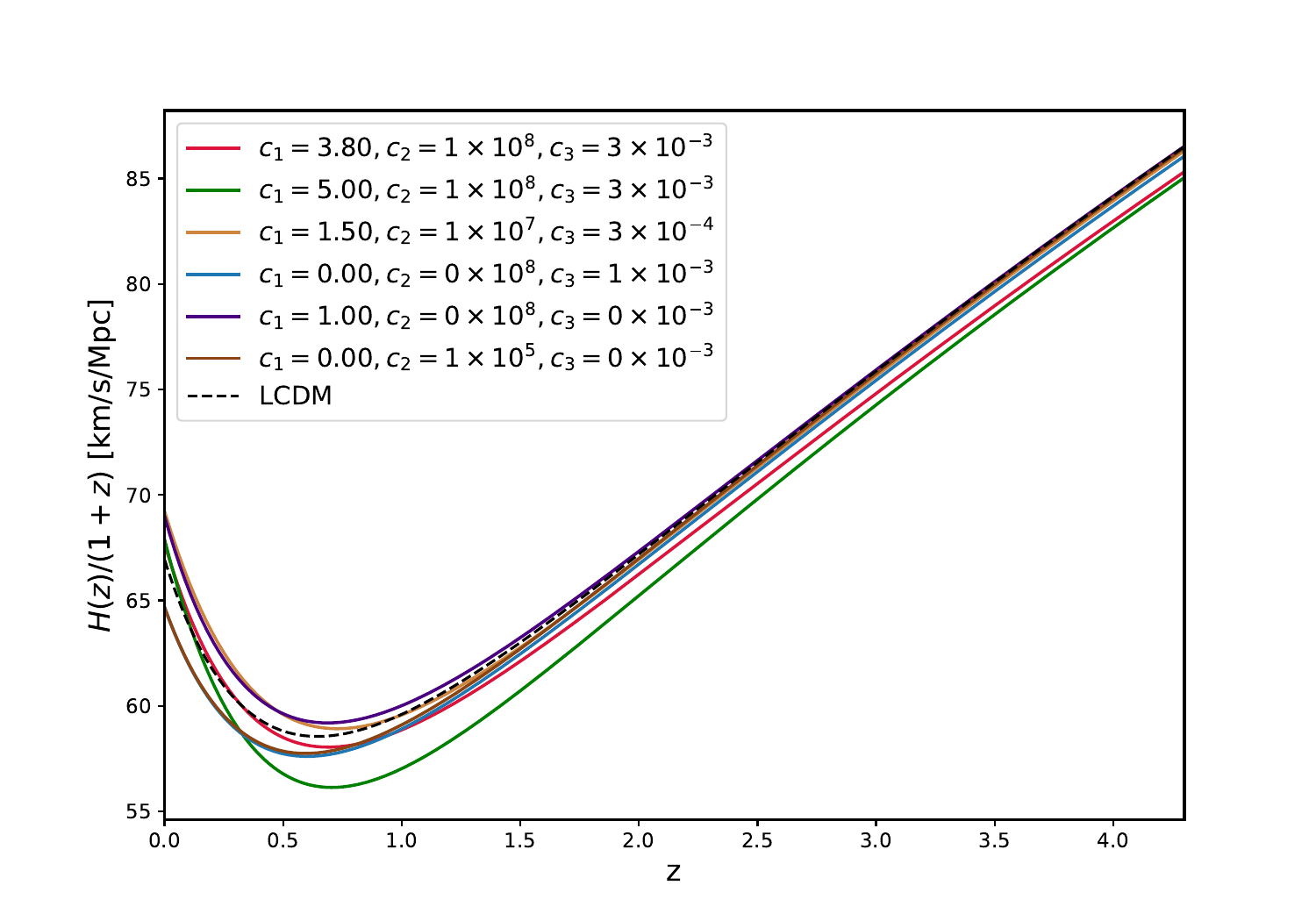}
    \caption{Evolution of Hubble parameter with redshift for different values of the model parameters (colored) and for $\Lambda$CDM (black dashed).}
    \label{fig:fig1}
\end{figure}

The background dynamics is governed by the homogeneous evolution of the energy density and pressure through the Friedmann equation. In our model, the scalar field serves as the dark energy component. Its effect on the background dynamics is most prominently visible in the evolution of the Hubble parameter and the dark energy equation of state. To maintain consistency with the physics of the early universe, we set initial conditions for the scalar field shortly after recombination, ensuring alignment with the minimal quintessence model, where dark energy density remains negligible. Consequently, the Hubble parameter initially follows the standard $\Lambda$CDM evolution and diverges at later times as the strength of various coupling parameters comes into play. Figure \ref{fig:fig1} illustrates the impact of these parameters on the Hubble expansion rate and thereby on the $H_0$, with all other cosmological parameters fixed to the Planck 2018 best-fit values \cite{Planck:2018vyg}.
\begin{figure}
    \centering
    \includegraphics[width=0.52\textwidth]{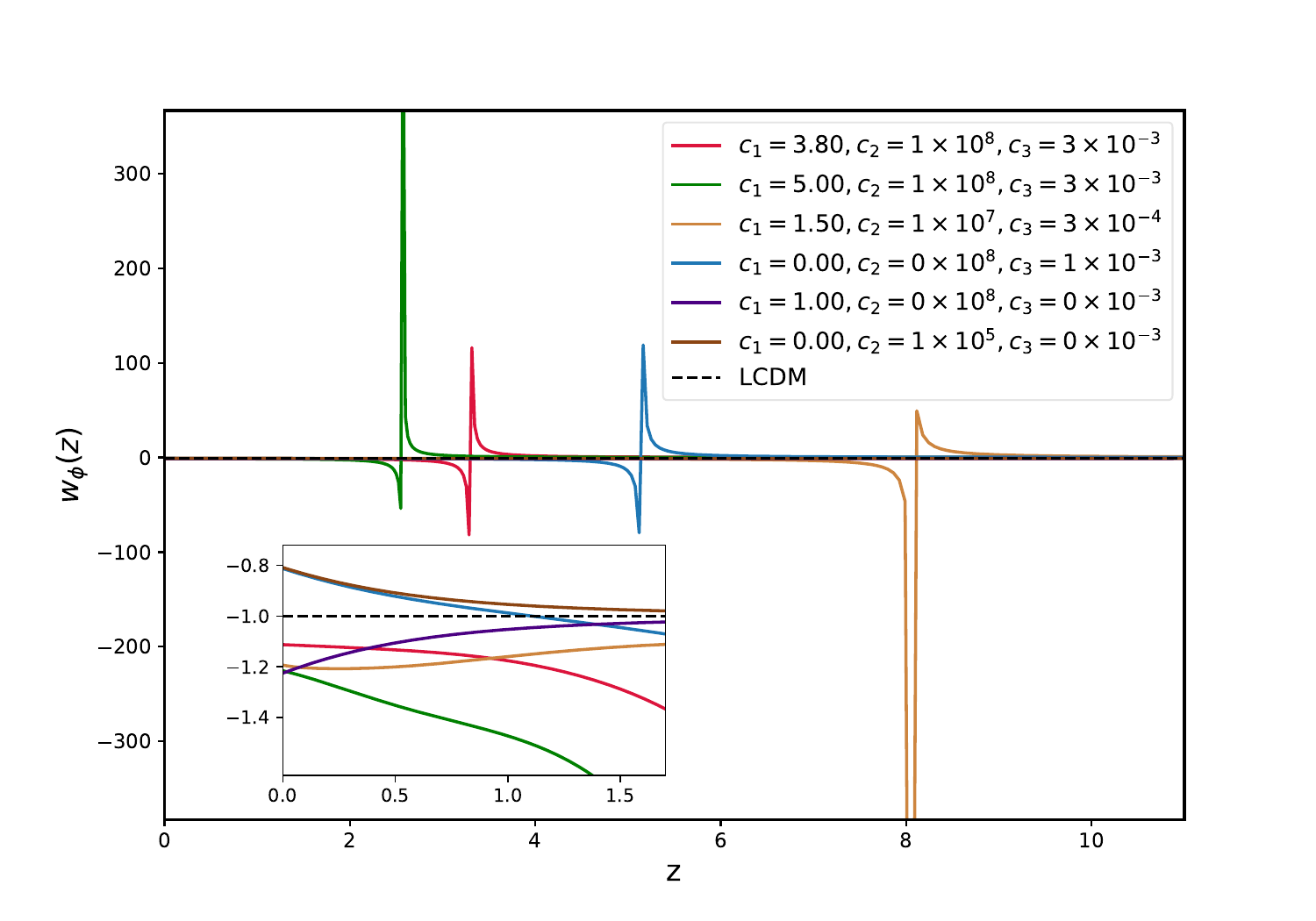}
    \caption{Evolution of dark energy equation of state with redshift for different values of the model parameters (colored) and for $\Lambda$CDM (black dashed). The inset plot shows the behavior at low redshifts.}
    \label{fig:fig2}
\end{figure}
To elucidate the specific behaviour of scalar field energy density and pressure at different epochs, we look into the evolution of its equation of state parameter  $w_\phi$ in Figure \ref{fig:fig2}. The inset plot of Figure \ref{fig:fig2} highlights the present-day equation of state. Self-interactions become significant at lower redshifts, as their dependence on the scalar field's kinetic term amplifies their impact only at late times. Notably, the parameter $c_1$ governs the equation of state at late times ($z<1$). As demonstrated in Figure \ref{fig:fig1}, increasing $c_1$ while holding $c_2$ and $c_3$ constant (depicted by red and green curves) leads to a faster expansion in comparison to the $\Lambda$CDM model. This effect is also seen in Figure \ref{fig:fig2}, where the present-day equation of state resides predominantly in the phantom regime, enhancing the expansion rate. Conversely, when $c_1=0$, the late-time equation of state shifts to the quintessence regime after briefly crossing the phantom divide, as shown by the blue and brown curves, leading to a slower expansion rate at $z<1$ compared to the standard model. The parameter $c_2$ plays a crucial role in ensuring $Q_s>0$, a condition that prevents ghost instabilities, as discussed in Appendix \ref{sec:appendix}.  Without a non-zero $c_2$, the background dynamics may remain stable, but $Q_s<0$, making the model physically unviable. To understand the effect of non-minimal coupling, we can look at the expression for the effective energy density obtained by using Eq. (3) in Eq. (A6) of Appendix \ref{sec:appendix},
\begin{equation}
    \rho_\phi= \frac{1}{2}\dot\phi^2+V(\phi)-6 c_3 \phi H^2 - 6 c_3 H \dot\phi - c_1 \dot\phi^2 + 3 c_2 H \dot\phi^3  
    \label{e:rho-ourmodel}
\end{equation}
For non-zero and positive values of $c_3$, energy density of the scalar field reaches negative values at high redshifts due to the dominance of the third term in Eq. (4). One straightforward consequence is that this reduces the expansion rate in comparison to $\Lambda$CDM during the period where energy density remains negative. Later on, as the universe expands and $H(z)$ decreases, other terms become dominant, leading to a positive energy density for the scalar field. Despite the negative values of $\rho_\phi$ at high redshifts, the total energy density of the universe remains positive, as the dark energy contribution is subdominant during those epochs. Consequently, the entire cosmological evolution remains free of instabilities.
However, the transition to a positive dark energy density is essential to explain the observed accelerated expansion at low redshifts. This transition from negative to positive energy density involves crossing zero, resulting in a singularity in the equation of state, which appears as a pronounced spike due to finite step size in numerical calculations, as shown in Figure \ref{fig:fig2}. This is physically acceptable, as both energy density and pressure contribute similarly to gravitational effects, and the absence of one does not inherently lead to any issues. Since $w$ denotes a derived quantity used for convenience to describe the behavior of dark energy, a singularity does not cause any unphysical behavior in the underlying dynamics. The precise location of these singularities is also influenced by the strength of the self-interaction terms, $c_1$ and $c_2$. Throughout this process, the scalar field consistently maintains negative pressure, allowing the equation of state to take on both positive and negative values across different cosmological epochs as the energy density evolves. In the absence of non-minimal coupling (i.e., for $c_3=0$), no singularities or negative energy densities occur, as depicted by the purple and brown curves. 
In summary, the dynamics reveal intriguing features, with self-interactions (\( c_1 \) and \( c_2 \)) primarily governing the late-time evolution of \( H(z) \), while the non-minimal coupling (\( c_3 \)) dictates the behavior of scalar field at high redshifts. It is the interplay between \( c_1 \), \( c_2 \), and \( c_3 \) that significantly influences deviations from \(\Lambda\)CDM, resulting in higher or lower \( H_0 \) values. However, the optimal values for these parameters, along with the cosmological parameters, which give a good fit to the observational data, are determined through detailed data analysis, as discussed in Sec. \ref{sec:fifth}.

\subsection{Perturbations}

\begin{figure}
    \centering
    \includegraphics[width=0.52\textwidth]{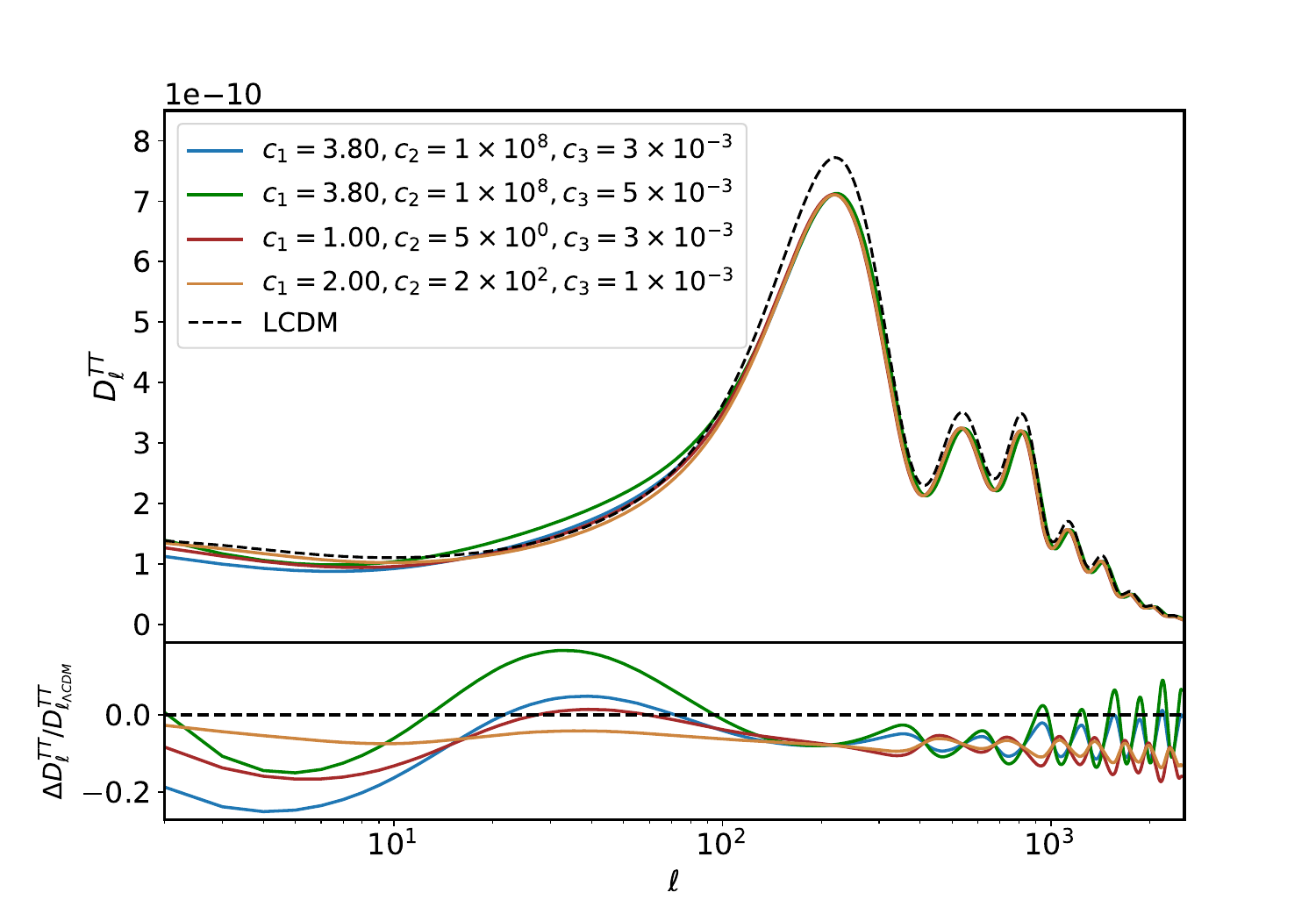}
    
    \caption{CMB temperature power spectrum (upper panel) and relative deviation (lower panel) from the $\Lambda$CDM (black dashed)  for different values of the model parameters (colored).}
    \label{fig:fig3}
\end{figure}
\begin{figure}
    
    \centering
    \includegraphics[width=0.52\textwidth]{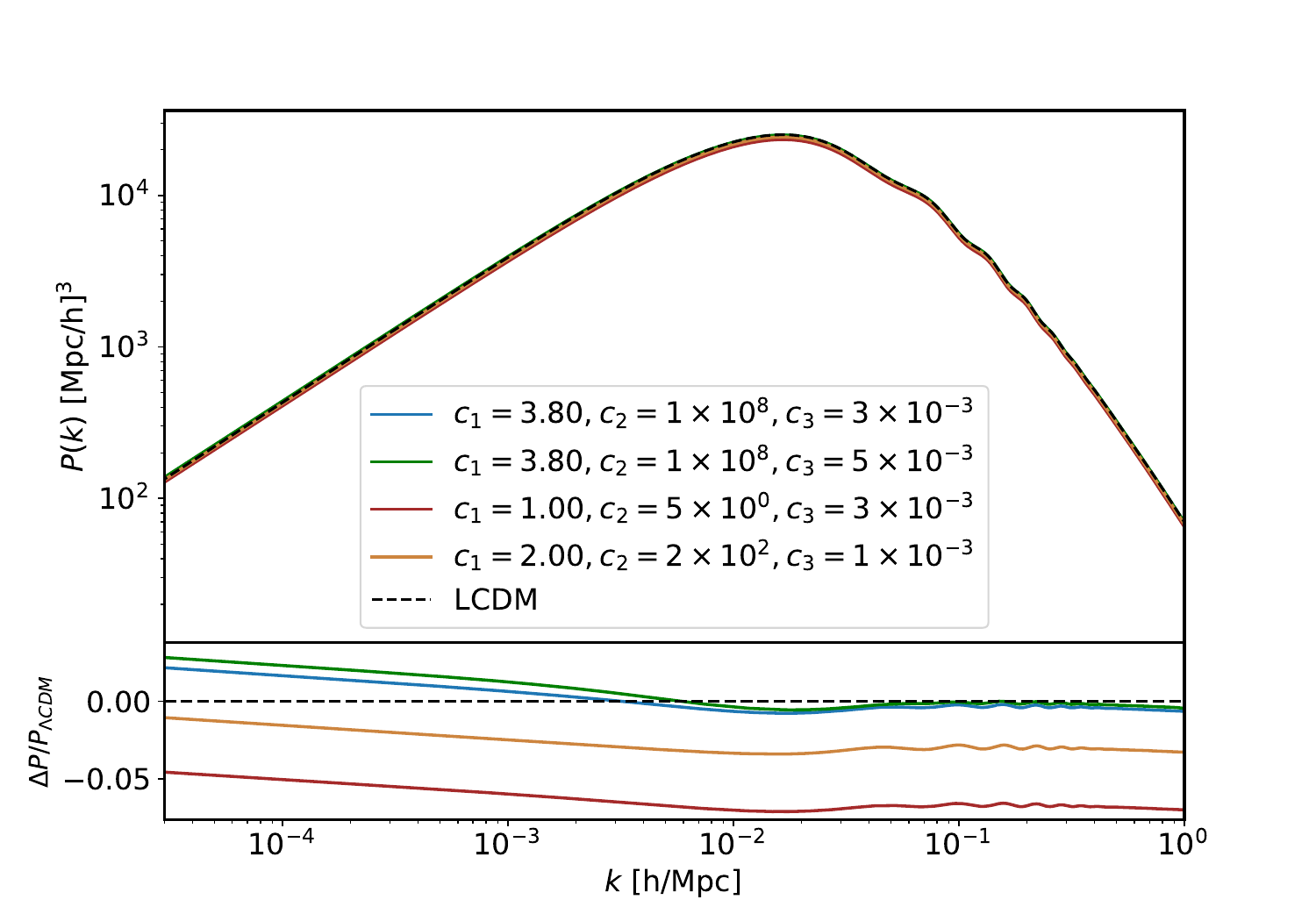}
    
    \caption{Matter power spectrum (upper panel) and relative deviation (lower panel) from the $\Lambda$CDM (black dashed) for different values of the model parameters (colored).}
    \label{fig:fig4}
\end{figure}

The most rigorous tests of cosmological models stem from comparing predicted and observed perturbations over the smooth background universe. In our framework, these perturbations are influenced through two primary channels: directly by fluctuations in the dark energy scalar field $\phi$ and indirectly through modifications in the background evolution. The background effect arises from changes in the evolution of the Hubble parameter $H(z)$, which can alter the distance to the last scattering surface. This has a significant impact on the Integrated Sachs-Wolfe (ISW) effect at low multipoles, as well as causing shifts in the acoustic peaks of the CMB power spectrum. The complex part of this analysis lies in the impact of perturbations in the dark energy field $\phi$.
In modified gravity theories like Horndeski, particularly those involving functions $G_3(\phi, X)$ and $G_4(\phi)$, there are notable modifications in the evolution of gravitational potentials compared to the standard cosmological model. These modifications prominently affect both the amplitude and the shape of the power spectrum on smaller scales. Figures \ref{fig:fig3} and \ref{fig:fig4} illustrate these effects on the CMB and matter power spectra for different model parameters, $c_1$, $c_2$, and $c_3$. In the lower panels, deviations from $\Lambda$CDM predictions are shown. An increase in the strength of non-minimal coupling, represented by $c_3$, leads to an enhanced amplitude of the peaks in the CMB power spectrum, while phase shifts on smaller scales largely result from self-interactions governed by parameters $c_1$ and $c_2$. Overall, certain parameter combinations can produce a suppression of the power at smaller scales, which in turn can lead to a reduction in the growth of structures.
\begin{figure}[h]
\centering
\includegraphics[width=0.48\textwidth]
{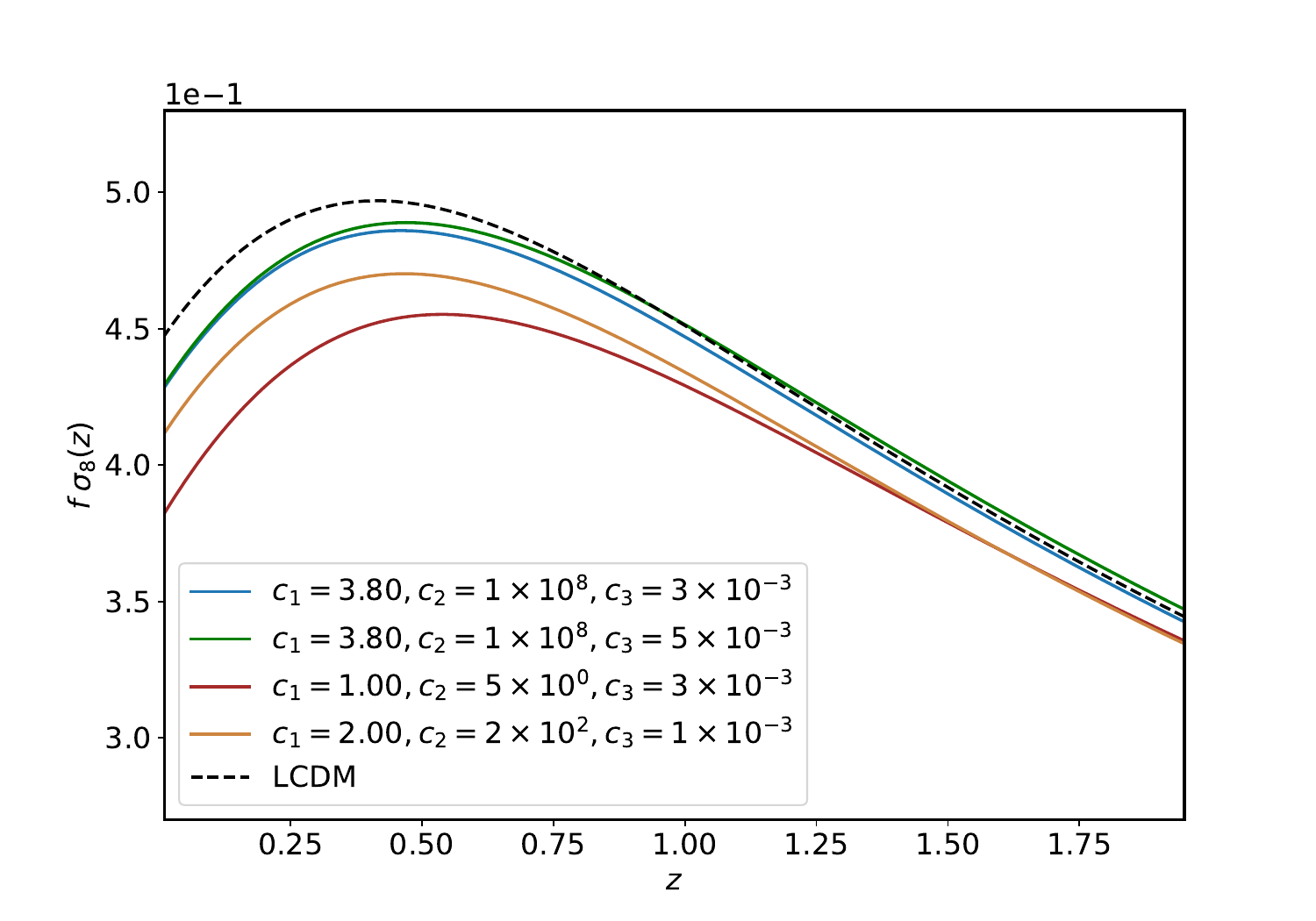}
\caption{The evolution of $f \sigma_8(z)$ for different values of the model parameters (colored). The black dashed line represents the $\Lambda$CDM model for reference.}
    \label{fig:fig5}
\end{figure}
The suppression of the matter power spectrum is particularly relevant to the growth tension. An alternative way to study the effect on the evolution of matter clustering is through $f \sigma_8(z)$. To understand this, consider the evolution of the matter density contrast $\delta=\delta \rho_{m}/\rho_m$ within the quasistatic approximation \cite{Matsumoto:2017qil},
\begin{equation}
    \Ddot{\delta}+2 H \dot\delta-4 \pi G_{\text{eff}}\rho_m \delta = 0
\end{equation}
For Horndeski theories, there could be an evolution in the effective gravitational coupling $G_\text{eff}$ \cite{Bellini:2014fua, Matsumoto:2017qil}, which leaves an imprint on the evolution of $\delta$ and subsequently on $f \sigma_8(z)$. Figure \ref{fig:fig5} illustrates the evolution of $f \sigma_8(z)$ for different parameter values. Notably, significant deviations from the $\Lambda$CDM prediction occur, particularly at low redshifts where the model exhibits a suppression in $f\sigma_8$. This feature aligns with the findings of \cite{Nguyen:2023fip} and is also useful in addressing the $S_8$ tension.
\section{Data Analysis}\label{sec:fifth}
As discussed in the previous section, our dynamical dark energy model exhibits several intriguing features, including phantom crossing and the possibility of negative dark energy at high redshifts. Additionally, the model demonstrates the potential for alleviating cosmological tensions within specific parameter ranges. To identify the regions of parameter space consistent with observational data, we perform a comprehensive statistical analysis. This involves integrating the Boltzmann solver {\tt hi\_class} with the statistical code {\tt Monte Python} \cite{Brinckmann:2018cvx, Audren:2012wb} for an efficient sampling from the multi-dimensional posterior distribution through the MCMC technique. \\

\subsection{Data and Priors}
We use the following datasets in the MCMC  analysis of our model:\\
\begin{itemize}
    \item \textbf{Planck18:} Low-$\ell$ TT, EE and high-$\ell$ TT, TE, EE likelihood from Planck 2018 \cite{Planck:2018vyg, Planck:2018lbu}.  
    \item \textbf{PantheonPlus:} Redshift-magnitude data compiled from 1701 light curves of 1550 spectroscopically confirmed Type-1a supernovae. This sample contains supernovae in the redshift range $z \in$ ($0.001,2.26$) \cite{Brout:2022vxf}.
    \item \textbf{BAO/$f\sigma_8$:} 
    BAO measurements from the BOSS DR12, 6dFGS, and SDSS DR7 survey along with $f\sigma_8$ measurements from BOSS DR12
    \cite{Beutler_2011, Ross:2014qpa, BOSS:2016wmc}.    
\end{itemize}
In our model, we have three additional parameters along with six $\Lambda$CDM  parameters, resulting in a total of nine parameters. We assign uniform priors to all the nine parameters. For the three model parameters $c_1$, $c_2$ and $c_3$, the sampling starts around zero, allowing them to take both positive and negative values without imposing strict bounds. This flexibility enables the MCMC sampling process to explore a broad range of parameter space comprehensively. A preference for non-zero values of these parameters would indicate potential evidence for dynamical dark energy effects or deviations from the standard cosmological model. To ensure consistency across analyses, we maintain identical priors for all datasets used in this study. Due to the increased complexity of this extended model with nine parameters, achieving good convergence in a full Bayesian analysis becomes computationally expensive. To ensure robust results, we continue the sampling process until the Gelman-Rubin convergence criterion \cite{Gelman:1992zz} is satisfied. Specifically, we set the convergence thresholds to $R-1 \lesssim 0.05$. 
This ensures that the Markov chains have sufficiently explored the parameter space and have converged to a stable posterior distribution.

\begin{figure}[htbp]
    \centering
    \includegraphics[width=0.48\textwidth]{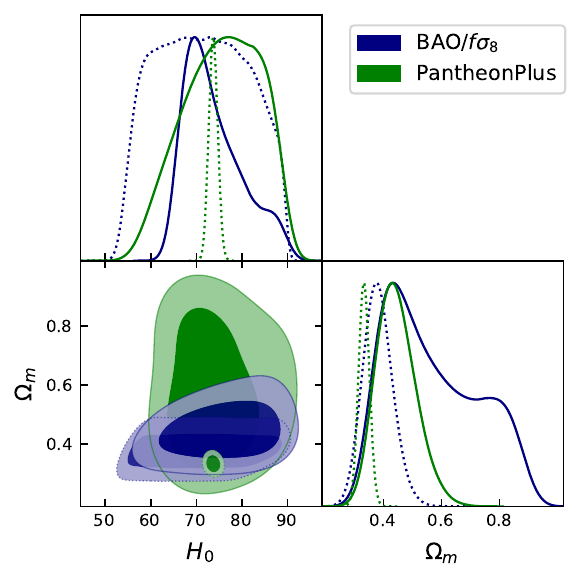}
    \caption{1D and 2D posterior distributions for $H_0$ and $\Omega_m$ obtained for the model (solid) and the $\Lambda$CDM (dashed), using PantheonPlus (green) and BAO/$f \sigma_8$ (navy) data. }
    \label{fig:fig6}
\end{figure}

\begin{figure}[htbp]
    \centering
    \includegraphics[width=0.48\textwidth]{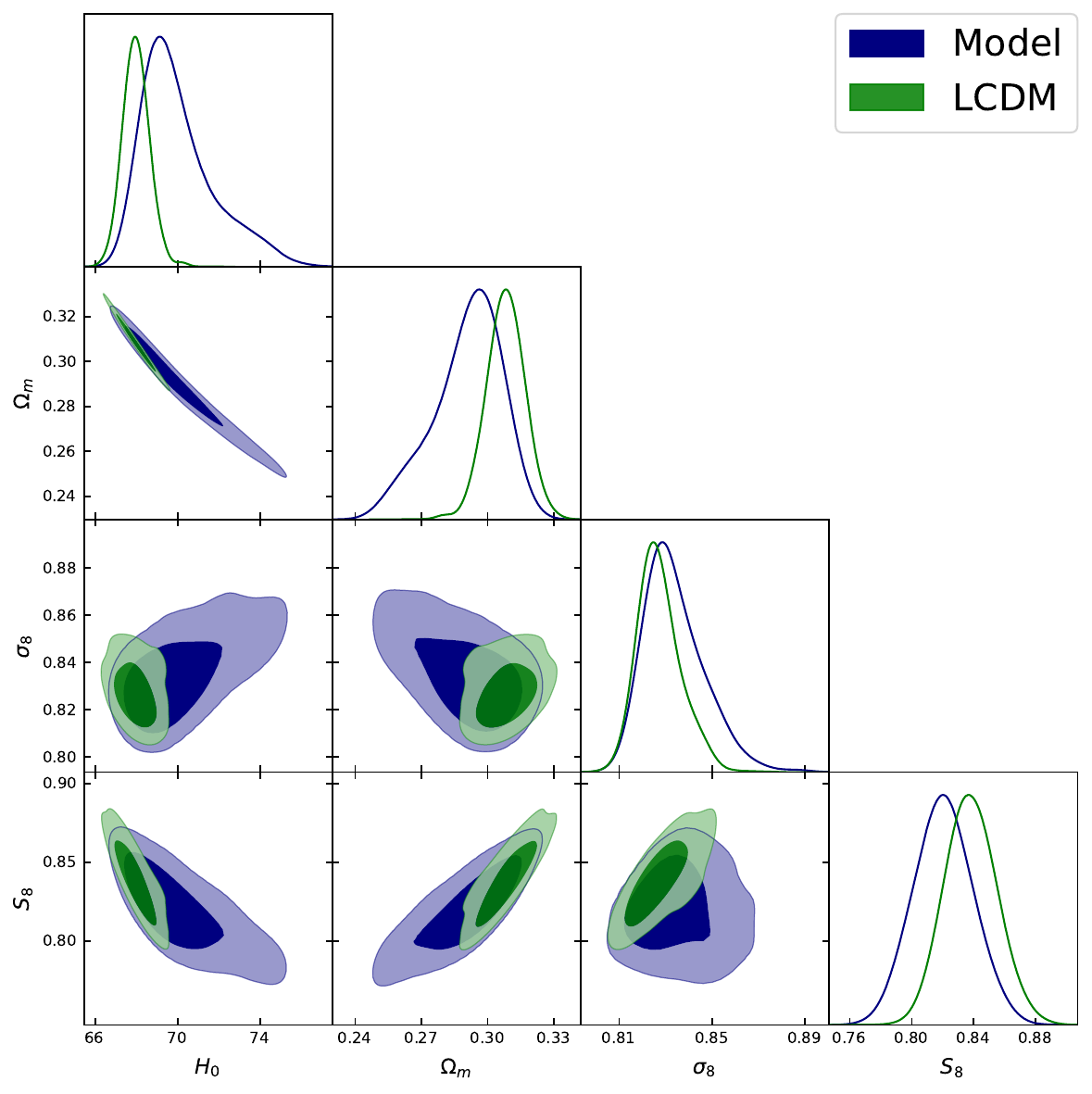}
    
    \caption{1D and 2D posterior distribution for the cosmological parameters for the model (navy) and the $\Lambda$CDM (green) obtained using Planck 2018 data. The model seems to prefer a higher value of $H_0$ and a lower value of $S_8$ parameter.}
    \label{fig:fig7}
\end{figure}

\begin{figure*}[htbp]
    \centering
    \includegraphics[width=\textwidth]{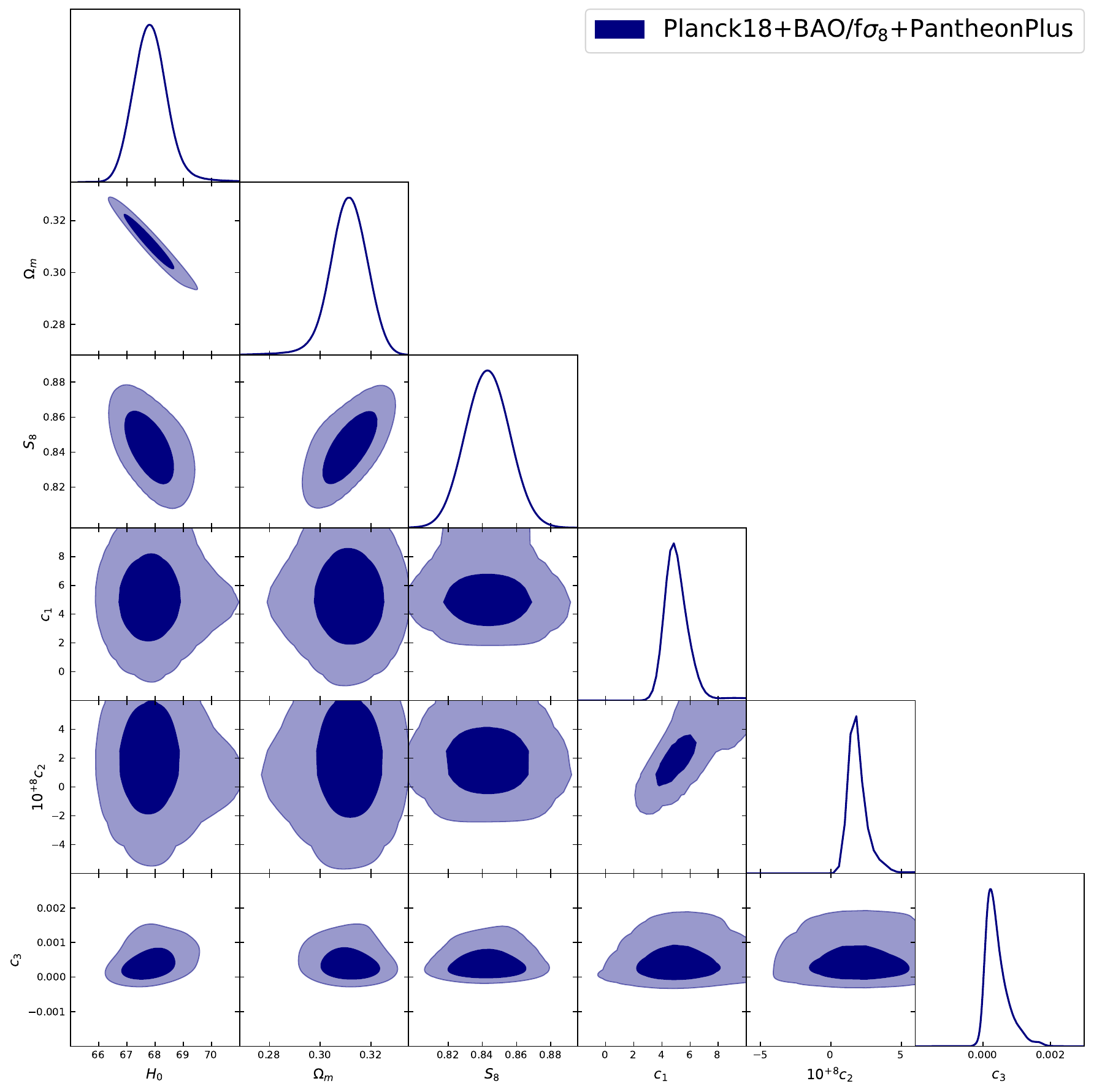}
    
    \caption{1D and 2D posterior distribution for a subset of cosmological parameters and the model parameters $c_1, c_2$ and $c_3$ obtained from the combined likelihood analysis using Planck18, BAO/$f\sigma_8$ and PantheonPlus data. All the three model parameters prefer a positive value within $1 \sigma$.}
    \label{fig:fig8}
\end{figure*}
 
\subsection{Results}
To better understand the implications of the model for cosmological observables, we begin our analysis by examining the aforementioned three datasets independently. Figures \ref{fig:fig6} and \ref{fig:fig7} show a comparative analysis of our model and the $\Lambda$CDM framework across these complementary datasets. The PantheonPlus and BAO/$f\sigma_8$ datasets primarily probe the late-time expansion history, providing constraints on key cosmological parameters: $\Omega_m$ and $H_0$. In Figure \ref{fig:fig6}, we show posterior distributions for $H_0$ and $\Omega_m$ from independent analyses of BAO/$f \sigma_8$ (navy) and PantheonPlus (green) data. For the BAO/$f\sigma_8$ dataset, the relatively sparse data points result in limited constraining power, allowing both $\Lambda$CDM (dotted lines) and our model (solid lines) to span a larger parameter space in the $H_0-\Omega_m$ plane. Similarly, the PantheonPlus dataset permits our model to explore a wider range of parameter space, including higher values of $H_0$. However, despite providing stronger constraints overall, PantheonPlus leaves our model less tightly constrained compared to $\Lambda$CDM, primarily due to the additional degrees of freedom in the model. The analysis with Planck18 data, shown in Figure \ref{fig:fig7}, offers additional insights. Unlike the background data, the CMB anisotropy offers significantly tighter constraints on the model parameters. When examining this dataset alone, our model (navy) predicts higher $H_0$ values and lower $\Omega_m$ compared to $\Lambda$CDM (green).  This shift results in a reduced $S_8$ value, indicating that our model could simultaneously address both the $H_0$ and $S_8$ tensions. However, the preferred parameter space of the model varies with the dataset, necessitating a joint analysis to achieve consistent and tighter constraints by leveraging the complementary strengths of all datasets.

In Figure \ref{fig:fig8}, we show the posterior distribution for the model parameters 
when the low redshift observations from Supernovae and BAO/$f\sigma_8$ are combined with the CMB power spectrum by Planck18. The corresponding mean values, along with $1\sigma$ error bounds and best-fit values, are summarized in Table \ref{tab:Table1}, alongside the results for the $\Lambda$CDM model for comparison. This analysis reveals that there is a preference for a positive, non-zero value of all the model parameters $c_1$, $c_2$, and $c_3$  within $1\sigma$ constraints. Of particular interest is the preference of nonzero and positive $c_3$ in the data. As discussed earlier, a positive $c_3$ implies a negative energy density for the dark energy field at high redshifts. This feature suggests that models with non-minimal coupling, characterized by a function $G_4(\phi)$, could offer compelling alternatives to GR. Additionally, the non-zero values of $c_1$ and $c_2$ show the preference for phantom-like behavior in the equation of state at low redshifts while avoiding ghost instabilities, thus maintaining theoretical consistency. However, the model does not outperform the standard $\Lambda$CDM scenario, with $\Delta\chi^2=0.6$, as shown in Table \ref{tab:Table1}.
Also, this combined analysis reveals that the model has difficulty accommodating significantly higher values of $H_0$ or lower values of $S_8$, limiting its ability to resolve cosmological tensions when considering all datasets together. These results contrast with individual dataset analyses, which suggested the model could alleviate these tensions. This highlights the importance of a unified background and perturbation analysis, as the final conclusions can change substantially.

\renewcommand{\arraystretch}{2.8}
\setlength{\tabcolsep}{2.2pt}

\begin{table*}[htbp]
\begin{center}
\renewcommand{\arraystretch}{1.8}
\setlength{\tabcolsep}{14pt}
\begin{tabular}{|c|c c|c c|}
 
 \hline
\multirow{2}{*}{Parameters} & \multicolumn{2}{c|}{Model} & \multicolumn{2}{c|}{$\Lambda$CDM} \\ 
\cline{2-5}
 & \multicolumn{1}{c}{Mean $\pm\; 1 \sigma$} & \multicolumn{1}{c|}{Best-fit} & \multicolumn{1}{c}{Mean $\pm \;1\sigma$} & \multicolumn{1}{c|}{Best-fit} \\ 
\hline

 $\Omega_bh^2$ & $0.02233 \pm 0.00015$ & $0.0223$ & $0.02242^{+0.00013}_{-0.00019}$ & $0.02242$ \\ 

 $\Omega_ch^2$ & $0.1208 \pm 0.0011$ & $0.1208$ & $0.1198^{+0.0014}_{-0.0009}$ & $0.1198$ \\

 $n_s$ & $0.9647^{+0.0038}_{-0.0043}$ & $0.9630$ & $0.9668^{+0.0036}_{-0.0052}$ & $0.9675$ \\

 $10^9A_s$ & $2.1093^{+0.028}_{-0.045}$ & $2.0988$ & $2.119^{+0.023}_{-0.057}$ & $2.1080$ \\

 $h$ & $0.6784^{+0.0052}_{-0.0060}$ & $0.6814$ & $0.6807^{+0.0039}_{-0.0066}$ & $0.6798$ \\

 $\tau$ & $0.0547^{+0.0065}_{-0.0097}$ & $0.0541$ & $0.0572^{+0.0063}_{-0.0120}$ & $0.0546$ \\

 \hline
 $c_1$ & $6.2716^{+0.42}_{-2.3}$ & $9.8487$ & $-$ & $-$ \\

 $10^{-8}c_2$ & $4.1704^{+1.7}_{-3.0}$ & $9.7245$ & $-$ & $-$ \\

 $c_3$ & $0.00042^{+0.00012}_{-0.00040}$ & $0.0004$ & $-$ & $-$ \\

 \hline
 $\Omega_m$ & $0.3111 \pm 0.0075$ & $0.3083$ & $0.3070^{+0.0084}_{-0.0054}$ & $0.3078$ \\

 $\sigma_8$ & $0.8278^{+0.0072}_{-0.0098}$ & $0.8273$ & $0.8273^{+0.0061}_{-0.0110}$ & $0.8256$ \\

 $S_8$ & $0.8430^{+0.0125}_{-0.0142} $ & $0.8386$ & $0.8369^{+0.0130}_{-0.0143}$ & $0.8362$ \\

 \hline
 \hline
 $\Delta\chi^2\equiv \chi^2_{Model}-\chi^2_{\Lambda \text{CDM}}$ & $-$ & $0.6$ & $-$ & $0.0$ \\


 \hline
 \hline
\end{tabular}
\vskip 10pt
\caption{The mean values with $1\sigma$ constraints and the best-fit values for the six $\Lambda$CDM parameters and three model parameters obtained from combined likelihood analysis using Planck18, BAO/$f\sigma_8$ and PantheonPlus data. The corresponding values for the $\Lambda$CDM model are also quoted for comparison.}
\label{tab:Table1}
\end{center}
\end{table*}

\section{Conclusion and Discussion}\label{sec:sixth}
The growing discrepancies between theoretical predictions and observational data, as well as inconsistencies among different datasets, have increasingly challenged the standard $\Lambda$CDM model \cite{Perivolaropoulos:2021jda}. Cosmological tensions, particularly in the measurements of $H_0$ and the $S_8$ parameters, suggest the potential need for physics beyond $\Lambda$CDM. 
Further challenges arise from observations of high-redshift galaxies by JWST, which reveal a population of massive galaxies at early epochs that are inconsistent with $\Lambda$CDM predictions \cite{2023Natur.616..266L, 2024Natur.635..311X, Arrabal_Haro_2023}. Explaining the formation of these galaxies necessitates non-standard conditions or modifications to the standard cosmological framework. Additionally, recent measurements of the expansion history using BAO features from the DESI survey combined with Supernova and CMB data point toward a preference for dynamical dark energy at a significance exceeding $2\sigma$. These combined observational concerns underscore the need for alternative models or extensions to the standard framework. One plausible direction is to go beyond GR and explore modified gravity scenarios. 
\par Motivated by recent advancements in cosmological background parameterizations and reconstructions that attempt to resolve various cosmological tensions, this work explores a dynamical dark energy model within the framework of Horndeski gravity. In \cite{Heisenberg:2022lob}, it has been highlighted that phantom crossing behaviour is also a necessary condition for late-time dark energy models to simultaneously alleviate $H_0$ and $S_8$ tensions. The significant advance of our work lies in deriving these novel features from the dynamics of a fundamental scalar field rather than relying on phenomenological parameterizations of dark energy properties. Additionally, the framework avoids common issues such as ghost or gradient instabilities, ensuring the theoretical consistency of the model while also exhibiting rich and interesting dynamics.  To constrain the parameter space and test the viability of the model, we conducted an extensive MCMC analysis using a comprehensive set of observational data. This included measurements from the CMB, BAO/$f\sigma_8$, and Supernovae. 
While the combined analysis of these datasets indicates that the model does not outperform the standard $\Lambda$CDM framework, giving a comparable fit to the data, there is still a notable preference for non-zero values of the model parameters within $1\sigma$ constraints. This suggests that such MG scenarios could provide compelling alternatives to GR.  Of particular interest is the preference for a positive value of the non-minimal coupling parameter $c_3$, which corresponds to negative dark energy densities at high redshifts. This feature may have important implications for structure formation and merit further investigation. 
Our analysis also indicates that in a specific region of the parameter space, the model possesses the necessary features to simultaneously address both $H_0$ and $S_8$ tensions. Interestingly, analyses with individual datasets indicated that the model could accommodate higher values of $H_0$ and lower values of $S_8$ simultaneously. However, when all datasets were combined in a joint analysis, particularly combining the low redshift probes like BAO/$f\sigma_8$ and Supernovae with CMB data, the model encountered strong constraints, limiting its ability to resolve the tensions.
This study highlights the critical importance of performing a thorough perturbation analysis when assessing such models proposed to resolve cosmological tensions. In scenarios with dynamical dark energy, the effects on cosmological perturbations—specifically those impacting the CMB anisotropy spectrum and the matter power spectrum—can be substantial. 
In particular, for models within the broader framework of modified gravity or beyond GR theories, the evolution of perturbations can deviate significantly from standard $\Lambda$CDM predictions, impacting observables like the ISW effect and growth rate of structures. These effects can significantly constrain the model even if it seems to give a good fit to the background evolution based on low-redshift observations alone. Consequently, incorporating CMB data with low-redshift datasets, such as BAO and Supernovae, is essential for robust parameter constraints. This ensures consistency across different scales, enhancing the reliability of the analysis, and provides a thorough evaluation of any model proposed as an alternative to the $\Lambda$CDM scenario. 
\par The central message of this work is that this class of models can introduce novel and complex dynamics, leading to profound implications for the overall cosmological evolution. These MG scenarios could thereby provide compelling alternatives to GR, paving the way for further observational tests. In particular, the preference for a positive value of the non-minimal coupling parameter $c_3$, which corresponds to negative dark energy densities, can significantly influence the growth of cosmic structures, as highlighted in recent studies \cite{Paraskevas:2023itu, Paraskevas:2024ytz}. This has important implications for understanding the formation of high-redshift galaxies observed by the James Webb Space Telescope (JWST), as explored in \cite{Adil:2023ara, Menci:2024rbq}. Furthermore, forthcoming full data releases from DESI, which already hint at dynamical dark energy in its initial findings, will be crucial for testing these models, particularly those featuring phantom crossing. 
Future large-scale structure (LSS) surveys, such as those conducted by the Vera Rubin Observatory, Euclid, and the Nancy Grace Roman Space Telescope, will provide critical data for probing these scenarios. These missions will provide precise measurements of galaxy clustering, weak lensing, and redshift-space distortions and will be instrumental in constraining the effects of non-minimal couplings and thereby constraining such class of modified gravity models. In addition, GW observations from current detectors like LIGO-Virgo-KAGRA and next-generation observatories such as the Einstein Telescope and LISA will play a transformative role. MG models predict distinctive GW signatures, including modifications to the GW luminosity distance and potential time delays between electromagnetic and GW signals \cite{Linder:2021pek}. These features provide a direct means to test Horndeski functions like $G_4(\phi)$, which govern non-minimal coupling, through their impact on cosmic evolution. The synergy between LSS surveys and GW observations will enable stringent tests of these models, offering a powerful framework to differentiate GR from MG scenarios. This will refine our understanding of the dark sector and offer profound insights into the fundamental nature of gravity and dark energy.

\section*{Acknowledgments} \label{sec:acknowledgments}
YT and UU would like to thank Shiv Sethi for insightful discussions on the impact of the Horndeski model on the power spectrum, and Elisa Ferreira for her input on data analysis.
YT also extends thanks to Emilio Bellini, Sandeep Haridasu, Matteo Viel, Tanvi Karwal, Ruchika, Anjan Sen and Shinji Mukohyama for their valuable feedback and discussions. 
YT also acknowledges Basundhara Ghosh for partial support through the DST Inspire Grant during this project. We acknowledge the use of high-performance computational facilities at the Supercomputer Education and Research Centre (SERC), and PTG Cluster at the Department of Physics, Indian Institute of Science, Bengaluru, India. YT and UU also thank Abhishek Tiwari for helping with the Planck likelihood installation on the PTG cluster.
RKJ acknowledges financial support from the IISc Research Awards 2024, SERB, Department of Science and Technology, GoI through the MATRICS grant~MTR/2022/000821 and the Indo-French Centre for the Promotion of Advanced 
Research (CEFIPRA) for support of the proposal 6704-4 under the Collaborative Scientific 
Research Programme.

\appendix
\section{Horndeski Theory} \label{sec:appendix}
In this appendix, we present the general expressions governing the evolution of the background and the dark energy scalar field within the Horndeski framework \cite{Horndeski:1974wa, Kobayashi:2011nu, Kobayashi:2019hrl}. As outlined in Sec. \ref{sec:second}, the full action for Horndeski gravity can be expressed as:
\begin{equation}
\mathcal{S} = \int \mathrm{d}^4x \sqrt{-g}\, \left(\sum_{i=2}^5 \mathcal{L}_{i}+\mathcal{L}_M\right)\,,
\label{e:horndeski_action}
\end{equation}

where $\mathcal{L}_M$ accounts for the matter (and radiation) components, and $\mathcal{L}_i$ are the different interaction terms in the Horndeski Lagrangian, as defined in Eq. (1). This framework can effectively model the dark energy phase of the universe with a scalar field $\phi$ coupled to gravity, either minimally or non-minimally, depending on the specific form of $G_i(\phi, X)$. 

In a flat FLRW background with 
$ds^2=-dt^2+a(t)^2 d\textbf{x}^2$, two Friedmann equations are given by \cite{DeFelice:2011bh},
\begin{widetext}
\begin{eqnarray}
    2X G_{2,X}-G_2+6X\dot\phi H G_{3,X}-2 X G_{3,\phi}-6H^2 G_4+24H^2 X(G_{4,X}+X G_{4,XX}-12H X \dot\phi G_{4,\phi X} \nonumber\\
    -6H \dot\phi G_{4,\phi}+2H^3 X \dot\phi(5G_{5,X}+2X G_{5,XX})-6H^2 X(3G_{5,\phi}+2X G_{5,\phi X})=-\rho_M
\label{e:friedmann_first}
\end{eqnarray}
    
\begin{eqnarray}
    & G_{2}-2X(G_{3,\phi}+\ddot\phi G_{3,X})+2(3H^2+2\dot H)G_4-12H^2 X G_{4,X}-4H \dot X G_{4,X}-8\dot H X G_{4,X}\nonumber\\& -8H X\dot X G_{4,XX} +2(\ddot \phi+2H \dot \phi)G_{4,\phi}+4X G_{4,\phi\phi}+4X(\ddot \phi-2H\dot \phi)G_{4,\phi X}-2X(2H^3 \dot \phi+ 2H \dot H \dot \phi +3H^2 \ddot\phi)G_{5,X}\nonumber\\ & -4H^2 X^2 \ddot\phi G_{5,XX}+ 4H X(\dot X-H X)G_{5,\phi X}+2[2(\dot H X+H \dot X)+3H^2 X]G_{5,\phi}+4H X \dot\phi G_{5,\phi \phi}=-p_M
\label{e:friedmann_second}
\end{eqnarray}
\end{widetext}

These equations can be simplified as follows \cite{Matsumoto:2017qil}:,

\begin{equation}
    3H^2= \kappa^2(\rho_{\phi}+\rho_M)
    \label{e:friedman-first}
\end{equation}
\begin{equation}
    -3H^2-2 \dot{H} = \kappa^2(p_\phi+p_M)
    \label{e:friedman-second}
\end{equation}

where $\rho_\phi$ and $p_\phi$ denote the effective energy density and pressure of the scalar field, respectively and are given as,

\begin{widetext}

\begin{eqnarray}
    \rho_\phi=2X G_{2,X}-G_2+6X\dot\phi H G_{3,X}-2 X G_{3,\phi}-6H^2 G_4+24H^2 X(G_{4,X}+X G_{4,XX}-12H X \dot\phi G_{4,\phi X} \nonumber\\
    -6H \dot\phi G_{4,\phi}+2H^3 X \dot\phi(5G_{5,X}+2X G_{5,XX})-6H^2 X(3G_{5,\phi}+2X G_{5,\phi X})+\frac{3 H^2}{\kappa^2}
    \label{e:rho-phi}
\end{eqnarray}

\begin{eqnarray}
     p_\phi=G_{2}-2X(G_{3,\phi}+\ddot\phi G_{3,X})+2(3H^2+2\dot H)G_4-12H^2 X G_{4,X}-4H \dot X G_{4,X} \nonumber\\  -8\dot H X G_{4,X}-8H X\dot X G_{4,XX} +2(\ddot \phi+2H \dot \phi)G_{4,\phi}+4X G_{4,\phi\phi}+4X(\ddot \phi-2H\dot \phi)G_{4,\phi X}\nonumber\\ -2X(2H^3 \dot \phi+ 2H \dot H \dot \phi +3H^2 \ddot\phi)G_{5,X} -4H^2 X^2 \ddot\phi G_{5,XX}+ 4H X(\dot X-H X)G_{5,\phi X} \nonumber\\ +2[2(\dot H X+H \dot X)+3H^2 X]G_{5,\phi}+4H X \dot\phi G_{5,\phi \phi} -\frac{1}{\kappa^2} (3H^2+ 2 \dot H)
\label{e:pressure-phi}
\end{eqnarray}
\end{widetext}
In the above equations, $\kappa^2=1/M_\text{Pl}^2$, and from now on, we set  $\kappa^2=1$.  Finally, the equation for the evolution of the scalar field is given by \cite{DeFelice:2011bh},

\begin{widetext}    
\begin{equation}
    \frac{1}{a^3}\frac{d}{dt}(a^3
    \mathcal{J})=\mathcal{P}_\phi
\label{e:phi_evolution}
\end{equation}
where,

\begin{eqnarray}
    \mathcal{J}=\dot\phi G_{2,X}+6H X G_{3,X}-2 \dot\phi G_{3,\phi}+6H^2 \dot\phi(G_{4,X}+2X G_{4,XX})-12H X G_{4,\phi X} \nonumber\\
    +2H^3 X(3G_{5,X}+2X G_{5,XX})-6H^2 \dot\phi(G_{5,\phi}+X G_{5,\phi X}),
\label{e:J}
\end{eqnarray}

\begin{eqnarray}
    \mathcal{P}_\phi=G_{2,\phi}-2X(G_{3,\phi\phi}+\ddot\phi G_{3,\phi X})+6(2H^2+\dot H)G_{4,\phi}+6H(\dot X+2H X)G_{4,\phi X}\nonumber\\
    -6H^2 X G_{5,\phi \phi}+2H^3 X \dot \phi G_{5,\phi X}.
\label{e:P_phi}
\end{eqnarray}

\end{widetext}
In the equations above, $\rho_M$ and $p_M$ represent the energy density and pressure of the matter (including radiation), which satisfy the continuity equation,
\begin{equation}
    \dot \rho_M+3 H \rho_M (1+w_M)=0,
\label{e:continuity}
\end{equation}
where $w_M = p_M / \rho_M$ is the equation of state parameter for the fluid. The complete background evolution of the Universe can be determined by solving this system of equations. However, due to the complex and non-standard structure of the Horndeski Lagrangian, these models often face issues related to the stability of perturbations, which can render the background evolution unsuitable. Thus, it is essential to monitor the theoretical parameters, even when the focus is not directly on perturbation evolution for a given dark energy model. Instabilities in the perturbation evolution are typically categorized into two types: Laplacian (gradient) instability and ghost instability. Laplacian instability occurs when the squared sound speed of perturbations becomes negative, causing unstable growth of perturbation modes on small scales. Ghost instability, on the other hand, arises when the kinetic term of the perturbations has a negative sign, indicating a problematic quantum vacuum. To ensure a physically viable theory, we will focus on parameter regimes that avoid these instabilities.

Using the standard linear cosmological perturbation theory, one can derive the second-order action for scalar perturbations in the Horndeski framework as shown in \cite{DeFelice:2011bh},
\begin{equation}
    \mathcal{S}_2=\int \mathrm{d}t \mathrm{d}^3 x a^3 \left[Q_S\left(\mathcal{\dot R}^2-\frac{c_S^2}{a^2}(\partial_i \mathcal{R})^2\right)\right],
\label{e:action_scalartensor}
\end{equation}
Here, $\mathcal{R}$ denotes the scalar curvature perturbation. The quantities $Q_S$ is given by:
\begin{equation}
    Q_S \equiv \frac{w_1\left(4 w_1 w_3+9 w_2^2\right)}{3 w_2^2}
\end{equation}

The parameter $c_S$ is the propagation speeds of scalar modes, defined as,

\begin{widetext}
\begin{eqnarray}
   c_S^2 \equiv \frac{3\left(2 w_1^2 w_2 H-w_2^2 w_4+4 w_1 w_2 \dot{w}_1-2 w_1^2 \dot{w}_2\right)-6 w_1^2\left(p_M+\rho_M\right)}{w_1\left(4 w_1 w_3+9 w_2^2\right)}
\label{e:cssqr}
\end{eqnarray}
where,
\begin{equation}
w_1 \equiv 2\left(G_4-2 X G_{4, X}\right)-2 X\left(G_{5, X} \dot{\phi} H-G_{5, \phi}\right),
\label{e:w1}
\end{equation}
\begin{eqnarray}
w_2 &\equiv & -2 G_{3, X} X \dot{\phi}+4 G_4 H-16 X^2 G_{4, X X} H+4\left(\dot{\phi} G_{4, \phi X}-4 H G_{4, X}\right) X+2 G_{4, \phi} \dot{\phi}\nonumber\\
& +& 8 X^2 H G_{5, \phi X}+2 H X\left(6 G_{5, \phi}-5 G_{5, X} \dot{\phi} H\right)-4 G_{5, X X} \dot{\phi} X^2 H^2,
\label{e:w2}
\end{eqnarray}
\begin{eqnarray}
w_3 &\equiv & 3 X\left(K_{, X}+2 X K_{, X X}\right)+6 X\left(3 X \dot{\phi} H G_{3, X X}-G_{3, \phi X} X-G_{3, \phi}+6 H \dot{\phi} G_{3, X}\right) \nonumber\\
& +&18 H\left(4 H X^3 G_{4, X X X}-H G_4-5 X \dot{\phi} G_{4, \phi X}-G_{4, \phi} \dot{\phi}+7 H G_{4, X} X+16 H X^2 G_{4, X X}-2 X^2 \dot{\phi} G_{4, \phi X X}\right) \nonumber\\
& +& 6 H^2 X\left(2 H \dot{\phi} G_{5, X X X} X^2-6 X^2 G_{5, \phi X X}+13 X H \dot{\phi} G_{5, X X}-27 G_{5, \phi X} X+15 H \dot{\phi} G_{5, X}-18 G_{5, \phi}\right), \nonumber\\
\label{e:w3}
\end{eqnarray}
\begin{equation}
   w_4 \equiv 2 G_4-2 X G_{5, \phi}-2 X G_{5, X} \ddot{\phi} .
   \label{e:w4}
\end{equation}
\end{widetext}
For a stable and consistent evolution of perturbations in this theoretical framework, the following conditions must be satisfied to avoid gradient and ghost instabilities, respectively,
\begin{eqnarray}
    c_S^2>0, \hspace{.5cm} Q_s>0.
    \label{e: instability_check}
\end{eqnarray}
Next, we present the expressions for $Q_s$ and $c_S^2$ in our framework. To derive these, we substitute Eq. (4) into the above conditions.
\begin{widetext}

\begin{equation}
\scriptsize
Q_s= \frac{
2 \left(\frac{1}{2} + c_3 \phi\right) 
\left[
9 \left(4 H \left(\frac{1}{2} + c_3 \phi\right) + 2 c_3 \dot{\phi} - c_2 \dot{\phi}^3\right)^2 
+ 8 \left(\frac{1}{2} + c_3 \phi\right) 
\left(
\frac{3 \dot{\phi}^2}{2} 
+ 18 H \left(-H \left(\frac{1}{2} + c_3 \phi\right) - c_3 \dot{\phi}\right) 
+ 3 \dot{\phi}^2 \left(-c_1 + 6 c_2 H \dot{\phi}\right)
\right)
\right]
}{
3 \left(4 H \left(\frac{1}{2} + c_3 \phi\right) + 2 c_3 \dot{\phi} - c_2 \dot{\phi}^3\right)^2}
\end{equation}

\begin{equation}
\scriptsize
c_s^2 = \frac{
-2 (1 + 2 c_3 \phi) 
\left(p_M + \rho_M + \dot{H} (2 + 4 c_3 \phi) + 2 c_3 \ddot{\phi}\right) 
+ 4 c_3 H (1 + 2 c_3 \phi) \dot{\phi} 
+ 6 (c_2 \ddot{\phi} + 2 c_3 (c_3 + c_2 \phi \ddot{\phi})) \dot{\phi}^2 
+ 2 c_2 H (1 + 2 c_3 \phi) \dot{\phi}^3 
- 4 c_2 c_3 \dot{\phi}^4 
- c_2^2 \dot{\phi}^6
}{
\dot{\phi}^2 
\left(
2 + 4 c_3 (3 c_3 + \phi) 
- 4 c_1 (1 + 2 c_3 \phi) 
+ 12 c_2 H \dot{\phi} 
+ 3 c_2 \dot{\phi} \left(8 c_3 H \phi - 4 c_3 \dot{\phi} + c_2 \dot{\phi}^3\right)
\right)}
\end{equation}
\normalsize
\end{widetext}

In our analysis, we ensure that the stability conditions given in Eq. (A19) are consistently satisfied. This imposes stringent constraints on the parameter space of $c_1$, $c_2$ and $c_3$. Notably, the parameter $c_2$ plays a pivotal role in maintaining $Q_s>0$, thus preventing ghost instabilities while allowing the equation of state $w_\phi$ to enter the phantom regime.
Another distinctive feature of Horndeski theories is the time evolution of the speed of tensor perturbations, given by \cite{DeFelice:2011bh, Tiwari:2022zzz},
\begin{equation}
   c_{T}^2= \frac{G_4-X G_{5,\phi}-X G_{5,X}\Ddot{\phi}}{G_4-2X G_{4,X}-X(G_{5,X} \dot\phi H-G_{5,\phi})}
   \label{e:cTsqr}
\end{equation}
where recent observations of gravitational waves (GWs) from the LIGO-VIRGO collaboration and their electromagnetic counterparts have placed stringent bounds on the speed of GWs as follows \cite{Gong:2017kim, LIGOScientific:2017zic},
\begin{align}
-3\times 10^{-15}<c_T-1<7\times10^{-16} \,\label{e:cTbound}
\end{align}
This constraint implies that a viable dark energy model must satisfy Eq A23 across all regimes.
However, in this work, due to the specific choices of the functions $G_i$s as indicated by Eq. (4), the speed of tensor perturbations remains unaffected. Specifically, by setting $G_5(\phi, X)=0$ and $G_4(X)=0$, we ensure that the propagation speed of GWs is always equal to the speed of light. Consequently, we do not include the equations for tensor perturbations in the following framework.

\bibliography{sections/references}

\end{document}